\pdfoutput=1
\documentclass[fleqn,usenatbib]{mnras}

\usepackage{newtxtext,newtxmath}

\usepackage[T1]{fontenc}

\DeclareRobustCommand{\VAN}[3]{#2}
\let\VANthebibliography\thebibliography
\def\thebibliography{\DeclareRobustCommand{\VAN}[3]{##3}\VANthebibliography}

\usepackage{graphicx}	
\usepackage{amsmath}	
\usepackage{amssymb}	

\title[Synchrotron self-Compton Emission in TeV GRB 190114C]{Immediate Afterglow Physical Characteristics and Broadband Spectra Evidence Synchrotron self-Compton Emission as the Reason for VHE Production in TeV GRB 190114C}

\author[A. C. Krishna]{
Aadi C Krishna$^{1}$\thanks{Work performed while at the Shiv Nadar School Noida, India}
\\
$^{1}$Yale University, New Haven, CT 06511, USA}

\date{Accepted XXX. Received YYY; in original form ZZZ}

\pubyear{2022}

\begin{document}
\label{firstpage}
\pagerange{\pageref{firstpage}--\pageref{lastpage}}
\maketitle

\begin{abstract}
Long GRB 190114C, identified on January 14th, 2019, was the first Gamma-ray Burst that substantially violated the defined 10 GeV energy limit of the Synchrotron model, with an observed emission between 0.2 - 1 TeV and a low redshift of z = 0.425. This paper analyzes its immediate afterglow broadband spectrum from 10$^{17}$ to 10$^{26}$ Hz based on observations by the Swift X-ray Telescope (XRT), Fermi Gamma-Ray Burst Monitor (GBM), Swift Burst Alert Telescope (BAT), Fermi Large Area Telescope (LAT), and Major Atmospheric Gamma Imaging Cherenkov Telescope (MAGIC). We first calculate the physical characteristics necessary to understand the conditions in the burst's emitting region, then conduct temporal and spectral analyses by deriving light curves and spectra using a chain polynomial best-fit in the context of the forward shock model in a homogeneous circumburst density. The Spectral Energy Distributions are found to be double-peaked for T$_0$ + 68-180s, and we show the distribution consists of a distinct Synchrotron component followed by an inverse Compton component explained by high-energy electrons up-scattering Synchrotron photons. We find our calculated Bulk Lorentz Factor = 351 sufficiently explains the peak of the inverse Compton component at sub-TeV energy levels in the immediate afterglow, and that the Comptonization of the burst proceeds in the Klein-Nishina regime. We conclude this further evidences Synchrotron self-Compton emission as the mechanism behind the production of Very-High-Energy photons in GRB 190114C.
\end{abstract}




\section{Introduction}

\large Gamma-ray Bursts (GRB) are intense pulses of $\gamma$-rays emerging from the launch of ultra-relativistic jets along the rotation axis of collapsing stars \citep{Woosley}. The resulting burst is the most energetic electromagnetic emission in the universe, characterized by an initial short-lived highly-variable prompt emission followed by a long-lived broadband afterglow resulting from the external shocks produced in the interaction of the expanding jet with the circumstellar environment \citep{Meszaros}.

The Gamma-ray Burst, GRB 190114C, was first observed and localized on January 14th, 2019, at 20:57:02.63 UT (T$_0$) by the Fermi Gamma Ray-Burst Detector (GBM). Its detection recorded the highest ever prompt emission between 0.2 and 1 teraelectronvolts \citep{Mirzoyan}. A significant occurrence in high-energy Astrophysics, it challenged the foundational understanding of the processes involving the production of Gamma-ray Bursts. 

The burst triggered radio, microwave, optical, X-ray, and $\gamma$-ray detectors in several telescopes around the world, including the Swift X-ray Telescope (XRT), Fermi Gamma-Ray Burst Monitor (GBM), Swift Burst Alert Telescope (BAT), Fermi Large Area Telescope (LAT), and Major Atmospheric Gamma Imaging Cherenkov Telescope (MAGIC), which began observing the field from T$_0$ + 50s \citep{Hamburg}. The Nordic Optical Telescope (NAT) measured the redshift to be relatively low with $z = 0.4245$ \citep{Castro}. 

GRB 190114C's prompt emission consisted of a first multi-peaked pulse lasting for 15s, followed by a weaker pulse from 15-25s after the initial trigger \citep{Hamburg}. The peak luminosity was found to be L$_{peak}$ = $1.67 \times 10^{53}$ erg/s and total radiated energy $E_{\gamma,iso}$ = $2.5 \times 10^{53}$ erg \citep{Hamburg, Frederiks}. The burst was especially bright and produced over 30,000 counts/second above background in the most illuminated Sodium Iodide detector \citep{Ajello}. The early observations illustrated a delayed high-energy emission (more than 40 MeV) in the first few seconds and a subsequent transition to a harder spectrum \citep{Minaev}.

This teraelectronvolt detection was the first to significantly cross the 10 GeV theoretical limit of the standard Synchrotron model, previously used to explain the mechanism behind Gamma-ray Bursts \citep{Nakar, Abdo, Kumar, Arimoto}. Synchrotron self-Compton (SSC) has long been the strongest contender to present a plausible explanation of emissions that cross the limit, beyond which photon energy losses become more efficient than Synchrotron radiation \citep{Meszaros, Sari, Guetta, Beniamini, Lemoine, Galli}. SSC involves the additional inverse Compton component where some Synchrotron photons are up-scattered upon collision with high-energy electrons and gain huge amounts of energy before being radiated away. 

This research aims to evaluate if Synchrotron self-Compton emission is the reason for the Very-High-Energy photons produced in GRB 190114C. First, to understand the conditions in the emitting region of the burst, we investigate the physical characteristics: the Bulk Lorentz Factor, Microphysical parameters, and Comptonization regime. Then, we use a chain polynomial best-fit model to analyze multi-wavelength light curves and Spectral Energy Distributions of forward-shock accelerated electrons under a broken-power law model during the immediate afterglow, from T$_0$ + 68s to T$_0$ + 180s. 


\section{Physical Characteristics}
\label{sec:physical} 

Determining the physical characteristics of the emitting region is the first step towards understanding the mechanism behind the burst's production. We analyze two physical characteristics: the Bulk Lorentz Factor and Microphysical parameters.

\subsection{Bulk Lorentz Factor}

Lorentz factors illuminate the change in comoving frame properties with respect to time. The explosion ejecta's initial Lorentz factor, defined as the Bulk Lorentz Factor, $\Gamma_0$, is essential to understand Gamma-ray Burst production \citep{Ghirlanda}. We calculate $\Gamma_0$ by adopting the model proposed by \citet{Nava} that depends on:

\begin{enumerate}
    \item $t_p$, time at the peak of the GeV light curve
    \item $E_k$, isotropic equivalent kinetic energy of the burst
    \item $n$, circumburst density of the medium
\end{enumerate}

$\Gamma_0$ differs depending on the surrounding medium, either an Interstellar (ISM) or Wind medium, and can be calculated using the formula:

\begin{equation}
    \Gamma_0 = \left[\left(\frac{(17-4s)(9-2s)3^{2-s}}{2^{10-2s}\pi(4-s)}\right)\left(\frac{E_k}{n_0m_pc^{5-s}}\right)\right]^\frac{1}{8-2s}t_{p,z}^{-\frac{3-s}{8-2s}}
\end{equation}

Where $s = 0$ corresponds to the ISM case and $s = 2$ to the Wind medium case, $t_{p,z} = t_p/1+z$, $z$ is the measured redshift, $m_p$ the mass of the proton, and $n_0$ the normalization of the circumburst density profile: $n(R)=n_0R^{-s}$. Since we assume the ISM case\footnote{An ISM case was previously assumed for GRB 190114C by \citet{Wang}. \citet{MAGIC} further showed that both ISM and Wind medium circumburst densities explain the detected emission. To better compare our Spectral Energy Distribution with previous models, we also assume the ISM case.}, $n(R) = n_0$. With the observed $E_k$ = $3 \times 10^{53}$ erg, $n_0$ = 0.5 cm$^3$, and $t_p$ = 10s, we find $\Gamma_0$ = 351.

\subsection{Microphysical Parameters}
In addition to the Bulk Lorentz Factor, two other parameters, $\epsilon_e$ and $\epsilon_B$, are necessary to consider, providing an outlook into the relative importance of the emission components \citep{Sari}. These parameters are a function of various microscopic physical processes occurring during and after the expansion of the relativistic shock wave in the fireball model. $\epsilon_e$ is the fraction of the total internal energy used in accelerating the electrons in the shock, and $\epsilon_B$ is the fraction of the total explosion energy of the shock wave used in amplifying magnetic fields \citep{Medvedev}.

For the calculation of the Microphysical parameters, we adopt the single energy electron population model suggested by \citet{Derishev}\footnote{\citet{Derishev} posit the model is applicable to other electron distributions as well, including the more probable power-law distribution.}, where $\epsilon_e$ depends on the electron’s Lorentz factor, $\gamma_e$, that has different values depending on the Comptonization regime: Klein-Nishina (KN) or Thomson. 

If the Comptonization occurs in the KN regime, $\gamma_e$ is found using:

\begin{equation}
E_{IC}\simeq\Gamma\gamma_{e,KN}m_ec^2 \,\,\,\,\,\Rightarrow\,\,\,\,\,\gamma_{e,KN}\simeq\frac{ E_{IC}}{\Gamma m_{e}c^2}
\end{equation}

Otherwise, for the Thomson regime, $\gamma_e$ is found using:

\begin{equation}
E_{IC}\simeq\Gamma\gamma_{e,Th}^4\frac{B}{B_{cr}}m_ec^2 \,\,\,\,\,\Rightarrow\,\,\,\,\,\gamma_{e,Th}\simeq\left(\frac{ E_{IC}}{\Gamma m_{e}c^2}\frac{B_{cr}}{B}\right)^\frac{1}{4}
\end{equation}

Where E$_{IC}$ is the observed peak energy of IC photons, B is the observed Magnetic field strength = 2.5 G, and B$_{cr}$ is the Schwinger field strength = 4.5 $\times$ 10$^{13}$ G  \citep{Derishev}. 

Subsequently, $\epsilon_e$ is calculated by:

\begin{equation}
{\epsilon_e=
    \frac{\xi_e\gamma_em_e}{\Gamma m_p}}
\end{equation}

Where $\xi_e$, the number of electrons per baryon, is 0.87 for the ISM and 0.5 for the Wind medium \citep{Derishev}. For the ISM case, we find $\epsilon_e = 0.39$ in the KN and $\epsilon_e = 0.15$ in the Thomson regime.

Next, $\epsilon_B$ is found using:

\begin{equation}
\epsilon_B=
    \begin{Bmatrix}
    1 \\
    2/3
    \end{Bmatrix}
    {\LARGE \frac{4\eta_{bol}L_x^{iso}t}{(1+\eta_{IC})yE_{tot}^{iso}}}\\
\end{equation}

With the observed $\eta_{bol}$ = 2 and $y = 0.25$, we find $\epsilon_B = 2.13 \times 10^{-3}$ at $t = 68s$, in the range of previous simulations by \citet{Bosnjak} that demonstrated comparatively low magnetic fields are needed to produce a TeV emission.

Next, to establish the relative importance of the emission components and understand the possibility of the efficient production of SSC, we calculate the Luminosity Ratio, $x$, which is the ratio of $L_{IC}$ to $L_{syn}$, the luminosity of the inverse Compton and Synchrotron component, respectively. The expression for $x$, as derived by \citet{Sari}, follows the limits:

\begin{equation}
{\large x=\begin{cases} 
      \frac{\eta\epsilon_e}{\epsilon_B} &  \frac{\eta\epsilon_e}{\epsilon_B}\ll 1 \\
       (\frac{\eta\epsilon_e}{\epsilon_B})^\frac{1}{2}& \frac{\eta\epsilon_e}{\epsilon_B}\gg 1
   \end{cases}}
\end{equation}

Where $\eta$, the fraction of radiated electron energy, is $(\gamma_c/\gamma_m)^{2-p}$ during the slow-cooling, and 1 during the fast-cooling regime. The burst was observed to be in the fast-cooling regime during the immediate afterglow till 4000s \citep{MAGIC}; thus, $\eta = 1$.

The limit $\eta\epsilon_e/\epsilon_B \ll 1$ corresponds to the domination of Synchrotron cooling over the total emission, and $\eta\epsilon_e/\epsilon_B \gg 1$ results in the efficient production of the inverse Compton component and its dominance over the cooling \citep{Zhang, Panaitescu, Sari}. The Luminosity Ratio $x$, calculated at T$_0$ + 68s, follows the limit $\eta\epsilon_e/\epsilon_B \gg 1$ for both the KN and Thompson regime, signifying the dominance of inverse Compton over the Synchrotron component and, therefore, acting as a necessary precondition for the efficient production of SSC emission in GRB 190114C. 

\section{Temporal Characteristics}

After establishing the physical conditions of the burst, we proceed to understand the nature of the afterglow and discern the emission components responsible for Very-High-Energy photon production. 

\begin{figure}
\centering{\includegraphics[width=0.4\textwidth]{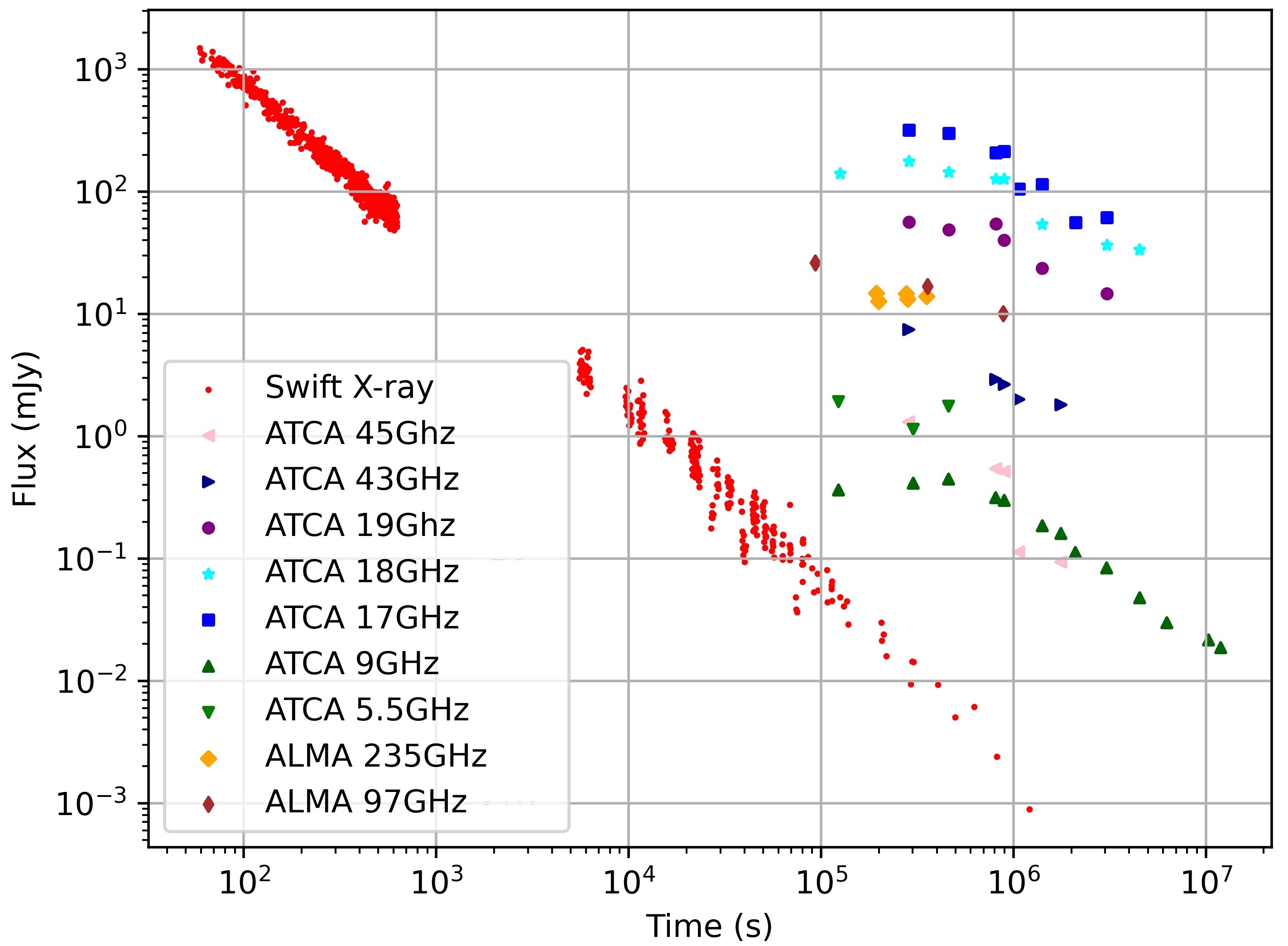}}
\caption{Multi-frequency Light Curves. The energy flux at different wavelengths observed by ALMA and ATCA are plotted alongside Swift from the early to late afterglow. The data and fluxes are multiplied by 10 for visibility.\label{fig1}}\vspace*{-3pt}
\end{figure}

In Figure \ref{fig1}, we plot light curves at different frequencies observed by the Swift X-ray Telescope (XRT), Australia Telescope Compact Array (ATCA), and Atacama Large Millimeter Array (ALMA). The graph shows transitions to emission components smoothly decaying as a power-law in time, supporting the interpretation of the presence of a primary Synchrotron component to explain the detection of low-energy photons. 

Next, to understand the spectral index $p$, the radiative flux per unit frequency, in Figure \ref{fig2} (Top Panel), we plot the early afterglow light curve for the Swift X-ray Telescope that observed GRB 190114C in the X-ray range of 0.3 - 10 keV. The flux density in the charted light curve is a power-law function of time represented by $F_x \propto t^{\alpha_x}$, where $\alpha_x$ is the power-law exponent and $\alpha$ is the photon index. We plot the logarithm equation in Figure \ref{fig2} (Bottom Panel) and find the flux to decay $\propto t^{-1.3969}$.

However, the light curve can decay in two possible cooling regimes: as $t^{(2-3p)/4}$ if $\nu_x > max (\nu_m,\nu_c)$ and as $t^{3(1-p)/4}$ if $\nu_m < \nu_x < \nu_c$, where $\nu_m$ is the characteristic Synchrotron frequency, $\nu_c$ is the cooling frequency, and $\nu_x$ is the frequency at a given point in time. The BAT and GBM observations show the light curve decay follows the bound $\nu_x > max (\nu_m,\nu_c)$ \citep{Ajello}; therefore, to calculate the spectral index, necessary to chart and evaluate the spectra in the next section, we use the equation $\alpha_x = (2-3p)/4$ \citep{Granot2} and find $p = 2.52$.

\begin{figure}
\centering{\includegraphics[width=0.4\textwidth]{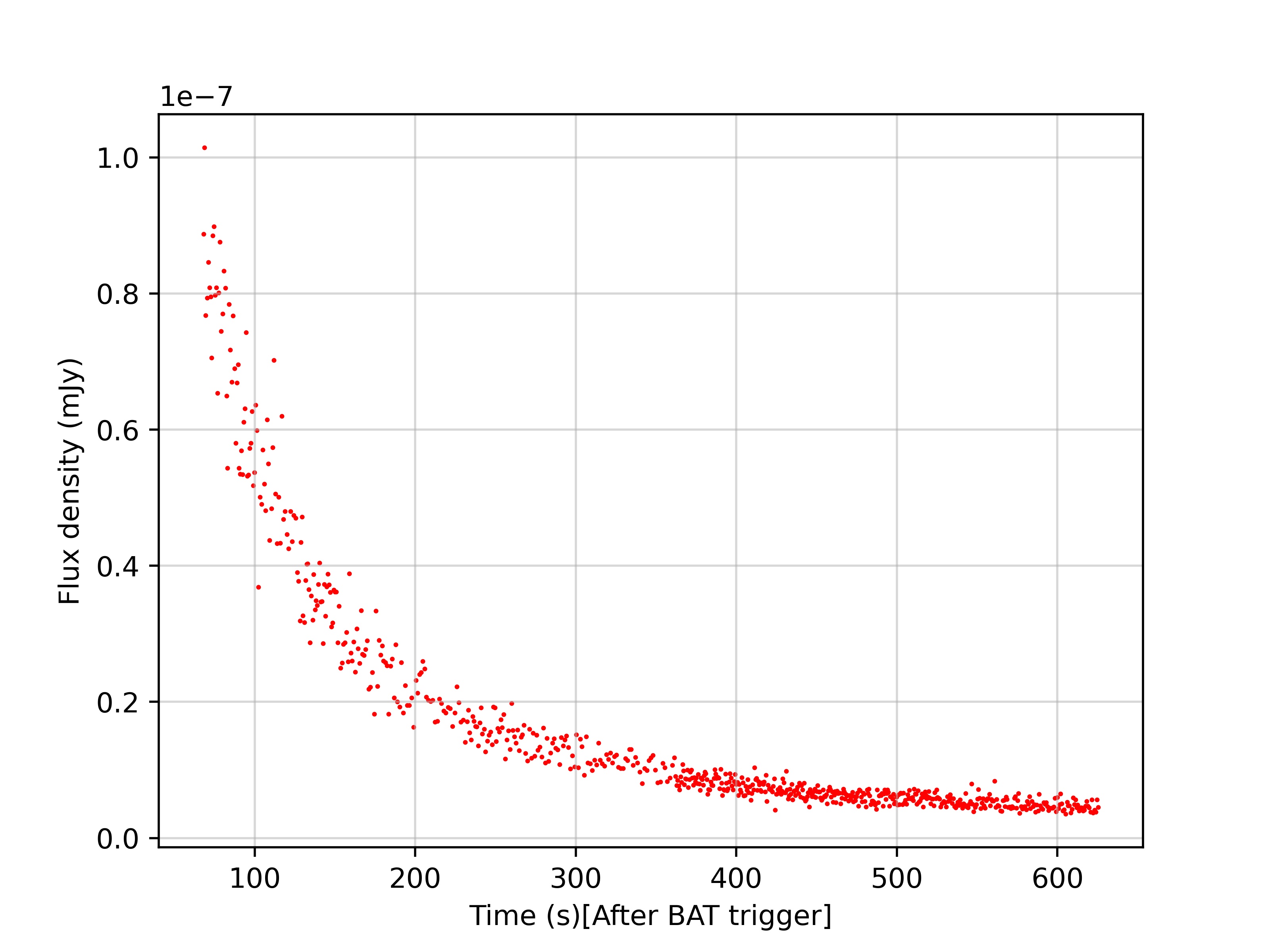}}
\end{figure}

\begin{figure}
\centering{\includegraphics[width=0.4\textwidth]{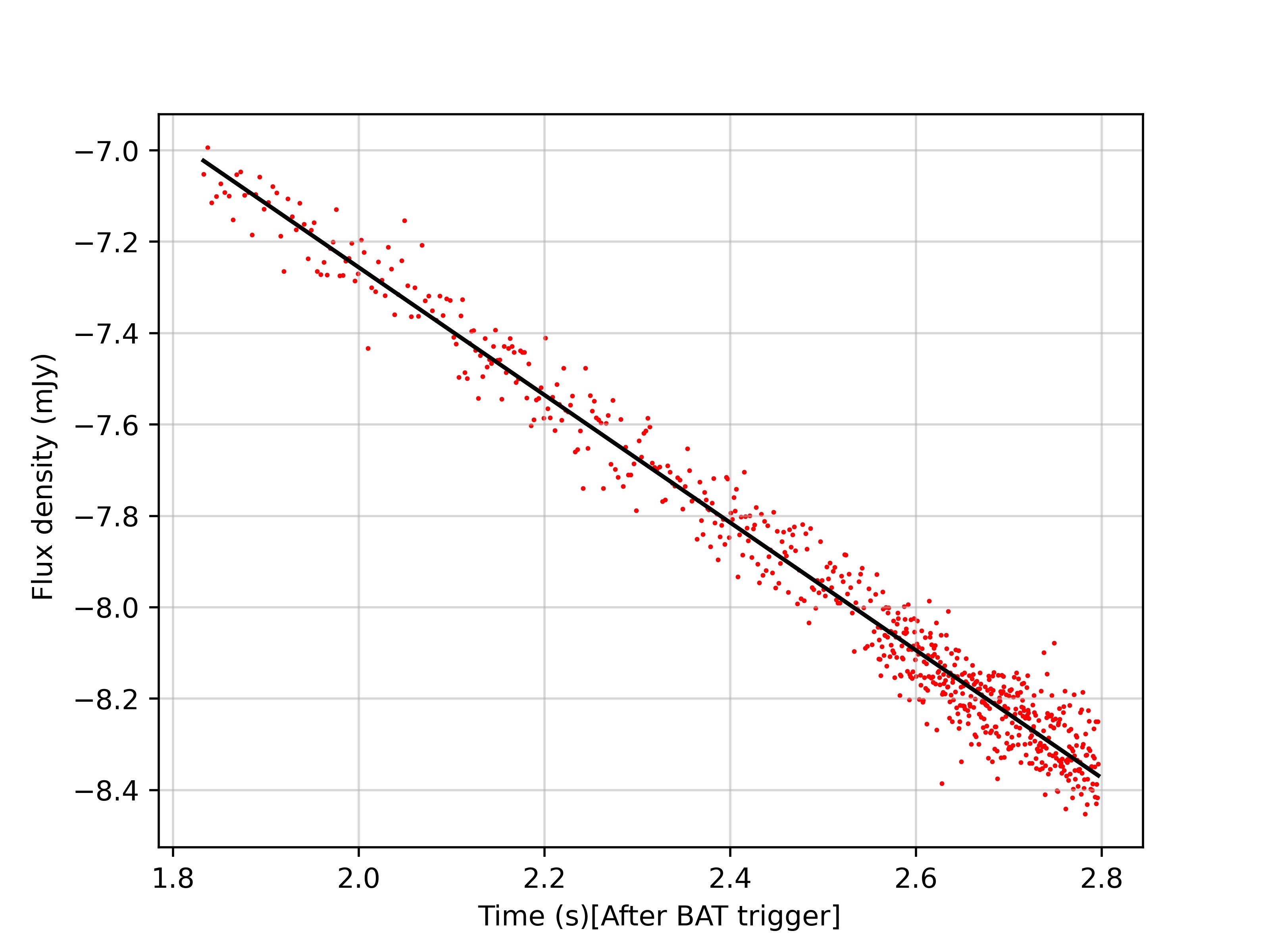}}
\caption{Top Panel: non-logarithmic Swift X-ray Flux vs Time. Bottom Panel: Swift X-ray Light Curve. The flux data was retrieved from the Swift XRT light curve repository and plotted with time after the BAT trigger: 20:57:03.19 UT on January 14th, 2019.\label{fig2}}\vspace*{-3pt}
\end{figure}

\section{Spectral Characteristics}
\label{sec:spectral}

The Spectral Energy Distributions (SED) of the radiation detected in the X-ray and $\gamma$-ray bands are then charted in Figure \ref{fig3} using a chain polynomial best-fit model. The model was built using a 5-degree \texttt{numpy.polyfit} chain polynomial from the \texttt{numpy} package library in Python computing the least-squares best-fit and restricted to the extreme data points.

The data used to model the SED were taken for two epochs, T$_0$ + 68s - 110s and T$_0$ + 110s - 180s, as the Swift-XRT and MAGIC observations started around this time \citep{Hamburg}, and the data of the complete broadband spectrum was available for these time bins. We only calculate and study forward-shock parameters since reverse shock parameters are prominent in optical and early X-ray frequencies, whereas Synchrotron self-Compton and its relevant spectra occur at late X-ray and $\gamma$-ray bands dominated by forward-shocks.

\begin{figure*}
\centering{\includegraphics[width=1\textwidth]{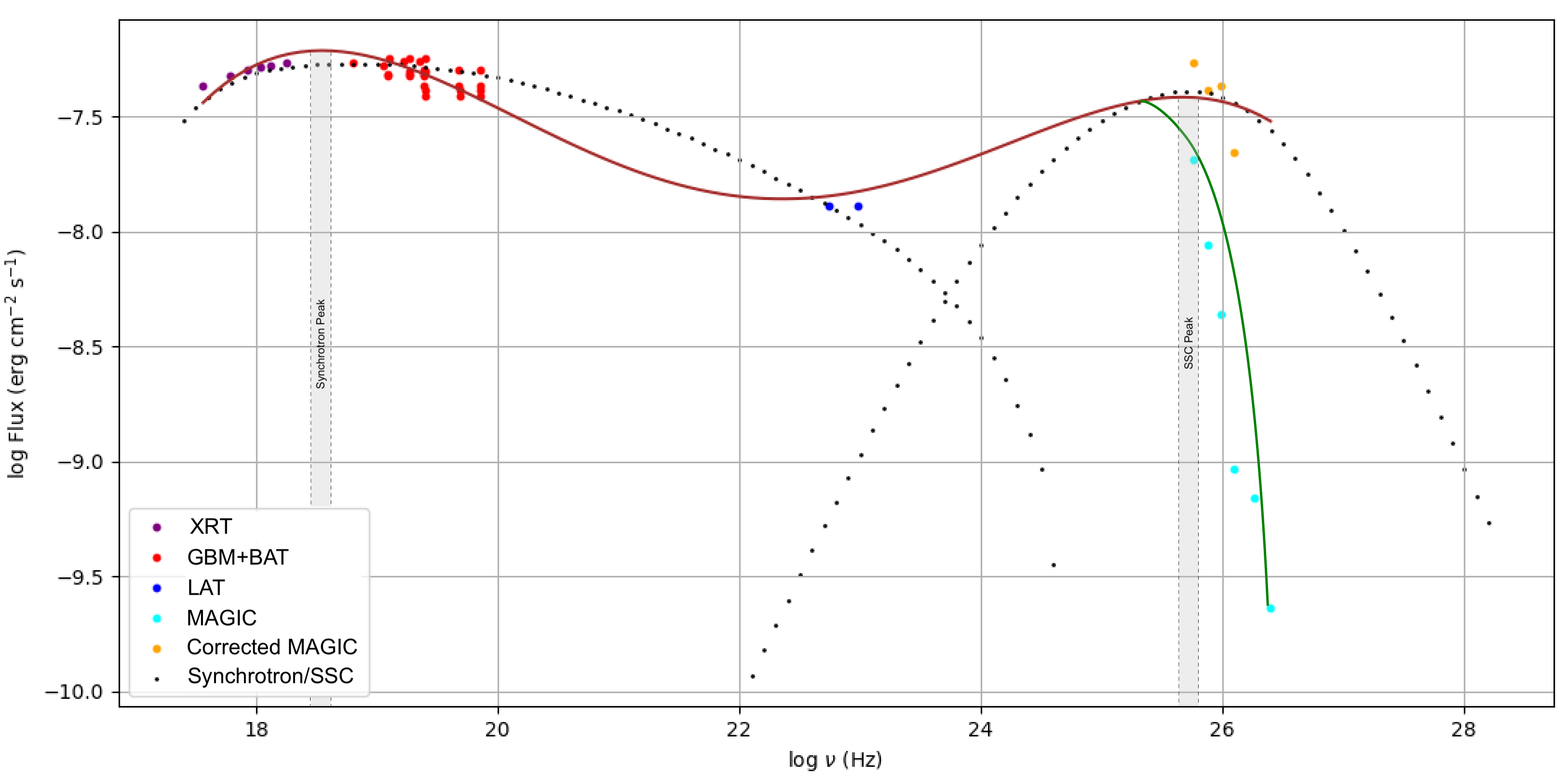}}
\end{figure*}

\begin{figure*}
\centering{\includegraphics[width=1\textwidth]{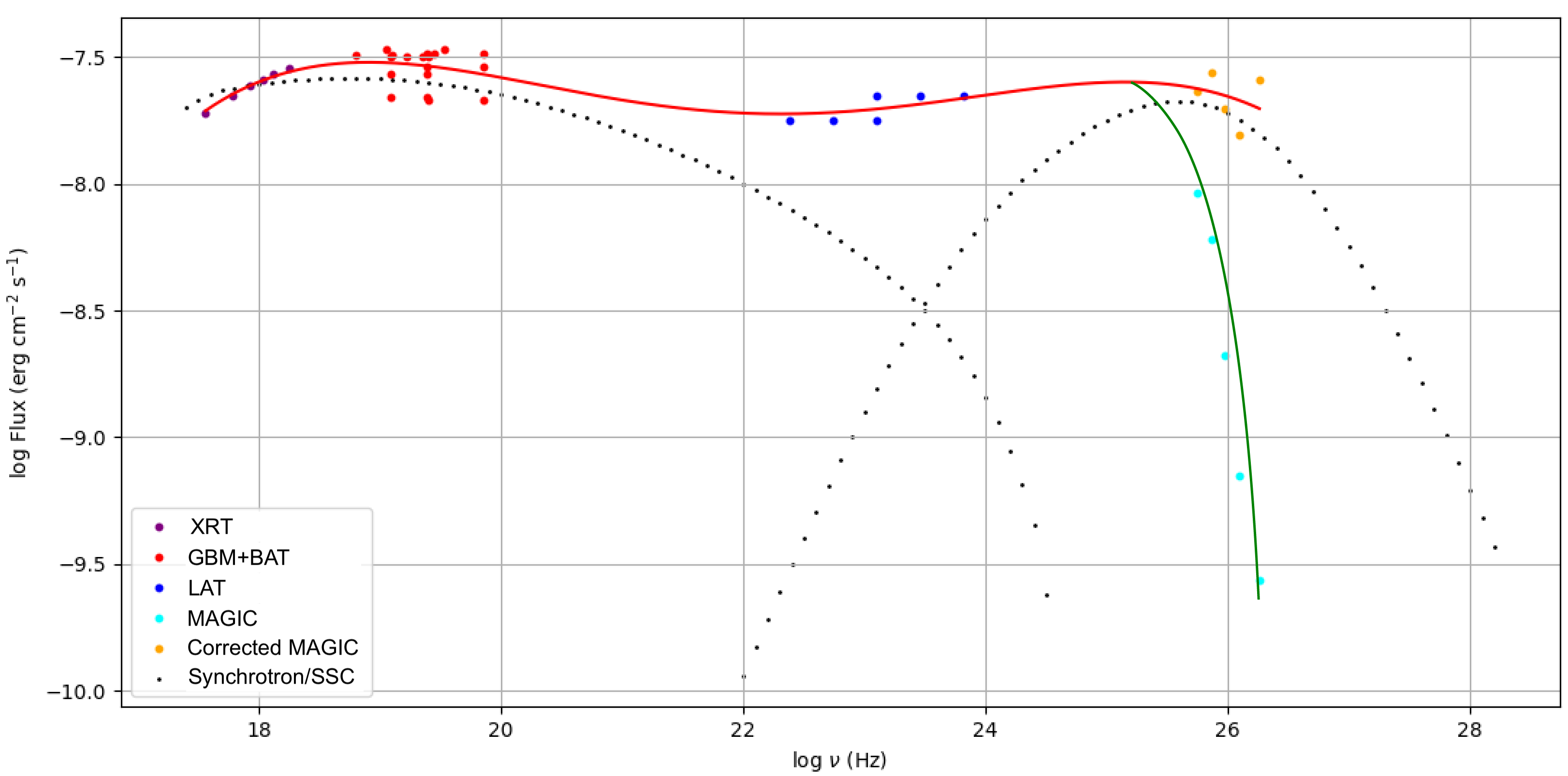}}
\caption{Modeling of the Spectral Energy Distributions in the time interval T$_0$ + 68-110s (Top Panel) and T$_0$ + 110-180s (Bottom Panel). Data from five telescopes are considered: XRT (Purple), GBM and BAT (Red), LAT (Dark Blue), and MAGIC (Aqua). The MAGIC data points corrected for attenuation due to EBL are represented by the Orange circles. The chain polynomial best-fit is shown by the Red line. The dotted line represents the modelled Synchrotron and SSC component of the afterglow by \citet{MAGIC} adopting the following parameters: s = 0, $\epsilon_e$ = 0.07, $\epsilon_B$ = $8 \times 10^{-5}$, p = 2.6, n$_0$ = 0.5 and E$_k$ = $8 \times 10^{53}$ erg.\label{fig3}}
\end{figure*}

The resultant spectra in Figure \ref{fig3} are double-peaked with a region that closely resembles the primary Synchrotron spectrum, followed by a higher energy component that can be explained by inverse Compton (IC) up-scattering of Synchrotron photons by high-energy electrons. In Figure \ref{fig3} (Top Panel), the Synchrotron peak energy is seen to be at 10 keV and SSC peak energy at 250 GeV with an area of superimposition in the middle. $\Gamma_0$ = 351 sufficiently explains the energy needed for the peak of the IC component to be at sub-TeV energy levels in the immediate afterglow. 

In the epoch T$_0$ + 110-180s (Figure \ref{fig3}, Bottom Panel), the burst loses energy, and the peak flux moves below $3 \times 10^7$ erg cm$^{-2}$ s$^{-1}$. The energy limit of Synchrotron emission is clearly visible as the chain polynomial best-fit line dips at early GeV energies. 

We find a higher than usual $\epsilon_B$ at T$_0$ + 68s, 2.13 $\times 10^{-3}$, implying a fast cooling regime for the inverse Compton electrons. The large flux of low-energy X-rays up till 180s in Figure \ref{fig3} further indicates, as suggested by \citet{Derishev}, that these IC electrons would remain in the fast cooling regime irrespective of the origin of the X-ray photons, providing favorable conditions for the production of SSC emission.

\subsection{Spectral Hardening \& Comptonization Regime}

Next, we analyze the spectra for spectral hardening at frequencies higher than the Synchrotron limit, a characteristic feature of SSC emission and essential to highlight the existence of another spectral component beyond the known Synchrotron radiation \citep{Nakar2}. In Figure \ref{fig3} (Top Panel), the spectrum can be seen to harden after the SSC transition frequency at GeV energy till the SSC peak. Simultaneously, low-energy X-rays are found to have a boosted flux, suggesting SSC already dominates the emission by T$_0$ + 68s.

The uncorrected observations of MAGIC further show hardening of the spectra after the second peak at sub-TeV energy (Green line, Figure \ref{fig3}). However, the corrected self-absorbed MAGIC values for photon attenuation due to the extragalactic background light (EBL) returns a soft spectrum (Yellow data points, Figure \ref{fig3}). 

One of the primary suspects for the observed soft spectrum of corrected MAGIC detection is Comptonization of photons in the KN regime \citep{Nakar2}. To confirm the Comptonization regime, we compare the relative values of $\gamma_e$ and $\gamma_{cr}$. As established by \citet{Derishev}, if $\gamma_e > \gamma_{cr}$, then the Comptonization is in the KN, otherwise in the Thomson regime. For IC photons to up-scatter, the energy of an electron, $\Gamma_0 \gamma_e m_e c^2$, must be greater than the energy of an IC photon, $E_{IC}$; implying that $E_{IC} \leq \Gamma_0 \gamma_e m_e c^2$. Since our value of $\Gamma_0$ is in the order of $10^2$, we can deduce that $\gamma_e \geq 10^4$. On the other hand, $\gamma_{cr}$ = $\frac{B_{cr}}{B}^{1/3} \,\,\,\,\,\Rightarrow\,\,\,\,\, \gamma_{cr}$ = $2.62 \times 10^{4}$. Since the approximate values of $\gamma_{e}$ and $\gamma_{cr}$, at the lower range of possible values for $\gamma_{e}$, are of the same order of magnitude, we cannot conclusively state which is larger. 

Therefore, we employ a second method that predicts the regime by comparing the values of the observed peak energy of the IC photons with that of the peak energy possible in the Thomson regime, $E_{IC}$ and $E_{IC}^{cr}$, respectively. If $E_{IC} \geq E_{IC}^{cr}$ then Comptonization occurs in the KN, otherwise in the Thomson regime. 

Using:

\begin{equation} 
E_{IC}^{cr} = \Gamma\left(\frac{B_{cr}}{B}\right)^{1/3}m_{e}c^2
\end{equation}

We find $E_{IC}^{cr}$ = 1.13 TeV. At the same time, $E_{IC}$ was observed up till at least 1 TeV, implying $E_{IC}^{cr} \approx E_{IC}$, but still not allowing us to deduce the regime conclusively. However, since we notice a visibly weaker and softer second component in the Spectral Energy Distribution (Figure \ref{fig3}), we can conclude that the Comptonization proceeds in the Klein-Nishina regime, and it is electrons radiating at the spectral peak in the KN regime that lead to the suppression of the SSC component.

Another reason behind the soft spectrum can be attributed to $\gamma\gamma$ absorption by X-ray photons \citep{Wang}. Although both KN suppression and $\gamma\gamma$ absorption become less efficient later in the afterglow, the spectral index in the sub-TeV band is expected to remain constant with time. This implies that the SSC peak would eventually cross the sub-TeV and reach the LAT energy band. 

However, a burst capable of emitting sub-TeV photons, as detected by MAGIC, signifies that the fireball must be optically thin ($\tau \leq 2$). This contradicts the higher value of optical depth ($\tau>2$) and opacity resulting from the capture and annihilation of Very-High-Energy photons by lower energy Synchrotron photons through electron-positron pair production. This "compactness problem" can be resolved by a value of the Bulk Lorentz Factor $\approx 10^2$ \citep{Baring}. We find the Bulk Lorentz Factor, $\Gamma_0$ = 351, satisfies the constraint, explaining the escape and detection of TeV photons from the ultra-relativistic emitting region after the production of Synchrotron self-Compton emission in the burst.

\section{Conclusion}

GRB 190114C provided the first unequivocal detection of TeV photons, enabling a detailed discussion of Synchrotron self-Compton emission beyond the theory essential for analyzing Very-High-Energy emissions in the future. We studied data spanning 10 orders of magnitude from 10$^{17}$ to 10$^{26}$ Hz in the immediate afterglow of the burst for a deeper look into the production of Very-High-Energy photons. The first part of the research focused on understanding the Bulk Lorentz Factor and Microphysical parameters in the burst's afterglow. These were found to present the necessary conditions for the efficient production of Synchrotron self-Compton emission with the Luminosity Ratio $x$ limit $\eta\epsilon_e/\epsilon_B \gg 1$ at T$_0$ + 68s and $\Gamma_0$ = 351.

In the second part, the temporal and spectral analyses of light curves and Spectral Energy Distributions showed the existence of another spectral component beyond the known Synchrotron emission. The research employed a chain polynomial best-fit model to chart the broadband spectra for T$_0$ + 68-110s and 110-180s and found it to be double-peaked. The second peak was seen to consist of an additional evidently delayed and brighter inverse Compton component with higher energy, which, however, remained weak due to Comptonization in the Klein-Nishina regime. The second component was distinguishable after the LAT energy range as the Synchrotron component transitioned to the Synchrotron self-Compton at GeV energies, and the spectrum hardened. We conclude that the physical characteristics and spectra of the immediate afterglow together evidence that Synchrotron self-Compton emission is the reason for the production of TeV photons in propagating forward shocks of long GRBs with low redshift such as GRB 190114C. 

\section*{Acknowledgements}

I would like to thank Vatsal Parmar (Ramnarain Ruia College, Mumbai) for his support in charting the figures and Team Tessaract at the Society for Space Education Research and Development for their discussions. 

\section*{Data Availability}

The data used to chart the Spectral Energy Distributions was collected from \citet{MAGIC} and can be accessed here: https://www.nature.com/articles/s41586-019-1754-6



\bibliographystyle{apalike}
\bibliography{example} 








\bsp	
\label{lastpage}
\end{document}